\documentclass[apj]{emulateapj}
\usepackage{apjfonts}
\usepackage{amsmath}
\usepackage{graphicx}
\usepackage{natbib}
\begin{document}

\title{Constraints on Natal Kicks in Galactic Double Neutron Star Systems}

\author{Tsing-Wai Wong, Bart Willems and Vassiliki Kalogera}

\affil{Department of Physics and Astronomy, Northwestern University, 2145 Sheridan Road, Evanston, IL 60208} 
\email{TsingWong2012@u.northwestern.edu, b-willems@northwestern.edu, vicky@northwestern.edu}

\begin{abstract}

Since the discovery of the first double neutron star (DNS) system in 1975 by Hulse and Taylor, there are currently 8 confirmed DNS in our galaxy. For every system, the masses of both neutron stars, the orbital semi-major axis and eccentricity are measured, and proper motion is known for half of the systems. Using the orbital parameters and kinematic information, if available, as constraints for \em all system\em, we investigate the immediate progenitor mass of the second-born neutron star and the magnitude of the supernova kick it received at birth, with the primary goal to understand the core collapse mechanism leading to neutron star formation. Compared to earlier studies, we use a novel method to address the uncertainty related to the unknown radial velocity of the observed systems. For PSR B1534+12 and PSR B1913+16, the kick magnitudes are 150 - 270 km/s and 190 - 450 km/s (with 95\% confidence) respectively, and the progenitor masses of the 2nd born neutron stars are 1.3 - 3.4 M$_\sun$ and 1.4 - 5.0 M$_\sun$ (95\%), respectively. These suggest that the 2nd born neutron star was formed by an iron core collapse supernova in both systems. For PSR J0737-3039, on the other hand, the kick magnitude is only 5 - 120 km/s (95\%), and the progenitor mass of the 2nd born neutron star is 1.3 - 1.9 M$_\sun$ (95\%). Because of the relatively low progenitor mass and kick magnitude, the formation of the 2nd born neutron star in PSR J0737-3039 is potentially connected to an electron capture supernova of a massive $O-Ne-Mg$ white dwarf. For the remaining 5 Galactic DNS, the kick magnitude ranges from several tens to several hundreds of km/s, and the progenitor mass of the 2nd formed neutron star can be as low as $\sim$1.5 M$_\sun$, or as high as $\sim$8 M$_\sun$. Therefore in these systems, it is not clear which type of supernova is more likely to form the 2nd neutron star.

\end{abstract}

\keywords{binaries: close --- pulsars: individual(PSR B1534+12, PSR B1913+16, PSR J0737-3039, PSR J1518+4904, PSR J1756-2251, PSR J1811-1736, PSR J1829+2456, PSR J1906+0746) --- stars: evolution --- stars: kinematics --- stars: neutron --- supernovae: general }

\section{Introduction}

Over the years, computation modeling and hydrodynamic calculations of supernova explosions due to core-collapse of massive stars have improved our theoretical understanding of how compact objects form. At present, two types of supernovae are actively discussed in the literature:
(i) After iron is synthesized within a massive star, the core completes its final nuclear burning state. It then becomes degenerate and keeps growing in mass due to silicon shell burning outside the core. Eventually, it crosses the Chandrasekhar mass limit and ensues collapse. When the core reaches nuclear density, it bounces back and explodes the star as a supernova. (Bethe 1990; Mezzacappa 2005; Woosley \& Bloom 2006; Kotake et al 2006; Janka et al. 2007)
(ii) On the other hand, if the core is not massive enough to ignite neon, nuclear burning within the core stops. The core becomes degenerate due to cooling by neutrino emission. Because of carbon shell burning outside the core, the mass of the core approaches the Chandrasekhar limit. When the core density and temperature rise above a certain threshold, electron capture by nuclei (such as $^{20}Ne$, $^{20}F$, $^{24}Na$, and $^{24}Mg$) takes place, decreasing degenerate electron pressure support. As a result, the core contracts rapidly and eventually bounces back, driving a supernova explosion. This scenario is referred to as electron capture supernova (ECS) (Miyaji et al. 1980; Nomoto 1987). Although the general idea of ECS has been established for several decades,  the mechanism of how electron capture occurs in these relatively less massive cores is not yet settled. 

Regardless of the supernova mechanism, it is expected that some asymmetries develop during the collapse, imparting a recoil kick to the nascent neutron star (see Harrison et al. 1993; Frail et al. 1994; Lyne \& Loimer 1994). The observed proper motions of 233 Galactic pulsars show that there is a class of neutron stars receiving high recoil velocities at birth (Hobbs et al. 2005). On the other hand, Pfahl et al. (2002) studied the orbital parameters of low eccentric Be/X-ray binaries, and pointed out that these neutron stars could only have received low kick velocities (< 50 km/s) at the time of formation. There is thus evidence for the existence of two different neutron star formation mechanisms (van den Heuvel 2004, 2007), potentially distinguishable by the magnitude of the natal kick of the neutron star.

Currently, there are 8 DNS observed in our galaxy. Their orbital parameters are summarized in table 1. The magnitude of the natal kick and the progenitor mass of the second-born neutron star (NS2) in these systems, except PSR J1906+0746, have been studied previously by Willems et. al. (2004, 2006), Thorsett et. al. (2005), Stairs et al. (2006), and Wang et al. (2006). In their study, Wang et al. (2006) studied the formation of \em all \em DNS except PSR J1906+0746, and derived constraints on the natal kick magnitude and progenitor mass of NS2 by studying the orbital dynamics of asymmetric supernova explosions, but not their kinematic history. The other authors focused on PSR B1534+12, PSR B1913+16, and PSR J0737-3039 and used both orbital dynamics of asymmetric supernova explosions and the measured proper motion of these systems to constrain the natal kick magnitude and progenitor mass of NS2. The authors traced the motion of the systems in the Galaxy backwards in time and assumed each crossing of the binary through the Galactic disk within the age of the DNS to be a possible birth site of the DNS. However, the backward calculation of the motion in the Galaxy requires the knowledge of the radial velocity which cannot be measured. Willems et al. (2004, 2006), Thorsett et al (2005) and Stairs et al. (2006) therefore considered possible radial velocities drawn from either a uniform or Gaussian radial velocity distribution.

In this paper, we avoid the assumption of a present-day radial velocity distribution by carrying out Monte Carlo simulations of the motion of the observed DNS in the Galaxy that is forward in time instead of backward. For this purpose, we distribute populations of DNS progenitors in the Galaxy according to a double exponential distribution function. Each system in the populations is assigned a velocity equal to the vector sum of the local Galactic rotational velocity and an isotropic kick velocity with a magnitude generated from the kick velocity distribution function, derived from the supernova orbital dynamics constraints for each of the observed systems. Compared to Willems et al.(2004), Thorsett et al. (2005), and Wang et al. (2006), we also relax the constraint that NS2's progenitor needs to be more massive than 2.1 M$_{\sun}$, which is a convectional limit for a helium star that explodes in a core collapse supernova. Instead, we restrict the progenitor to be more massive than NS2 only. Furthermore we study \em all \em eight known DNS in our galaxy, using the up-to-date observational parameters listed in Table 1.

The methodology of the adopted analysis is outlined in more detail in \S\,2, while results for the individual systems and comparison with earlier studies are discussed in \S\,3. In \S\,4, we summarize our results and discuss their implications for the supernova forming the second-born neutron star in the observed DNS.

\begin{deluxetable*}{@{}l@{}ccccc@{}ccc@{}c@{}c@{}c@{} }
\tablewidth{18.0 cm}
\tabletypesize{\footnotesize}
\tablecolumns{12}
\tablecaption{Parameters of the 8 known DNS in our galaxy}
\tablehead{ 
\colhead{System} & 
\colhead{$\alpha$ \tablenotemark{a} } & \colhead{$\delta$  \tablenotemark{b}} & \colhead{D \tablenotemark{c}} & \colhead{$\mu_\alpha$ \tablenotemark{d}} & \colhead{$\mu_\delta$ \tablenotemark{e}}  & 
\colhead{$\tau_c$ \tablenotemark{f}}  & \colhead{$M_1$ \tablenotemark{g}} & \colhead{$M_2$ \tablenotemark{h}} &\colhead{$A_{cur}$ \tablenotemark{i}}  & \colhead{$e_{cur}$ \tablenotemark{j}} &
\colhead{$\theta_{t}$ \tablenotemark{k}}
}
\startdata

PSR B1534+12\tablenotemark{r1} & 15 37 09.96 & 11 55 55.55 & 1.02 & 1.34(1) & -25.05(2) & 250 & 1.3332(10) & 1.3452(10) & 3.28 & 0.274 & 25(155) $\pm$ 3.8\\
PSR B1913+16\tablenotemark{r2} & 19 15 28.00 & 16 06 27.40 & 8.3(1.4) & -3.27(35) & -1.04(42) & 110 & 1.4408(3) & 1.3873(3) & 2.80 & 0.617 & 18(162) $\pm$ 6\\
PSR J0737-3039\tablenotemark{r3} & 07 37 51.25 &  -30 39 40.71 & 1.15 & -3.82(62) & 2.13(23) & 210 & 1.337(5) & 1.250(5) & 1.26 & 0.0878 & <15 \tablenotemark{l}\\
PSR J1518+4904\tablenotemark{r4} & 15 18 16.80 & 49 04 34.25 & 0.625 & -0.67(4) & -8.53(4) & 20000 & $0.72^{+0.51}_{-0.58}$ & $2.00^{+0.58}_{-0.51}$ & 24.7 & 0.249 & \nodata\\ 
PSR J1756-2251\tablenotemark{r5} & 17 56 46.63 & -22 51 59.40 & 2.5 &  -0.7(2) & \nodata & 443 & 1.312(17) & $1.258^{+0.018}_{-0.017}$ & 2.70 & 0.181& \nodata\\ 
PSR J1811-1736\tablenotemark{r6} & 18 11 55.03 & -17 36 37.70 & 6.0 &  \nodata & \nodata & 1830 & $1.62^{+0.22}_{-0.55}$ & $1.11^{+0.53}_{-0.15}$ & 40.7 & 0.828 & \nodata\\
PSR J1829+2456\tablenotemark{r7} & 18 29 34.60 & 24 56 19.00 & 1.2 &  \nodata & \nodata & 12400 & $1.14^{+0.28}_{-0.48}$ & $1.36^{+0.50}_{-0.17}$ & 6.36 & 0.139 & \nodata\\
PSR J1906+0746\tablenotemark{r8} & 19 06 48.67 & 07 46 28.60 & 5.4 &  \nodata & \nodata & 0.112 & 1.365(18) & 1.248(18) & 1.75 & 0.0853 & \nodata\\

\enddata

\tablenotetext{a}{Right ascension (J2000.0) (hours, minutes, and seconds)}
\tablenotetext{b}{Declination(J2000.0) (degrees, arcminutes, and arcseconds)}
\tablenotetext{c}{Distance (kpc)}
\tablenotetext{d}{Proper motion in R.A. (mas yr$^{-1}$)}
\tablenotetext{e}{Proper motion in dec. (mas yr$^{-1}$)}
\tablenotetext{f}{Characteristic age (Myr) }
\tablenotetext{g}{Mass of 1st born NS ($M_{\sun}$)}
\tablenotetext{h}{Mass of 2nd born NS ($M_{\sun}$)}
\tablenotetext{i}{Current semimajor axis ($R_{\sun}$)}
\tablenotetext{j}{Current orbital eccentricity}
\tablenotetext{k}{Misalignment angle between  pulsar spin axis and post-supernova orbital angular momentum axis (degree). The number in parenthesis correspond to another possible angle}
\tablenotetext{l}{Different data analysis model gives a different upper limit on this angle. See section 4.3 for details.}

\tablenotetext{r1}{Wolszczan 1991; Arzoumanian et al. 1999; Stairs et al. 2002; Konacki et al. 2003; Stairs et al. 2004 }
\tablenotetext{r2}{Hulse \& Taylor 1975; Taylor et al. 1976, 1979; Taylor \& Weisberg 1982, 1989; Damour \& Taylor 1991; Arzoumanian et al. 1999; Wex et al. 2000.}
\tablenotetext{r3}{Burgay et al. 2003; Lyne et al. 2004.; Lorimer et al. 2007; Ferdman 2008 et al. 2008; Breton 2009; Deller et al. 2009}
\tablenotetext{r4}{Thorsett \& Chakrabarty 1999; Janssen et al. 2008}
\tablenotetext{r5}{Faulkner et al. 2005; Ferdman 2008}
\tablenotetext{r6}{Lyne et al. 2000; Stairs 2004; Corongiu et al. 2007}
\tablenotetext{r7}{Champion et al. 2004, 2005; Stairs 2004; }
\tablenotetext{r8}{Lorimer et al. 2006, Kasian 2008, Stairs 2008}

\end{deluxetable*}

\section{Methods of Calculation}

Tauris $\&$ van den Heuvel (2004) reviewed the general scenario for forming DNS. It starts with a binary consisting of two massive ZAMS stars ( $\ga 10\, M_{\sun}$). The primary star, which is initially the more massive one, will leave the main sequence first and explode as the first supernova, in which it forms the first-born neutron star (NS1) in this binary. There may be mass transfer from the primary to the secondary star before that supernova takes place. As the secondary star leaves the main sequence, wind mass transfer from the secondary to NS1 can occur, leading to the formation of a X-ray binary. While NS1 is accreting mass from its companion, it is spun up or "recycled", and its spin axis is expected to align with the orbital angular momentum. Later on, the X-ray binary may end up in a common envelope phase, during which NS1 is engulfed by the extended envelope of its companion. The orbit shrinks and circularizes rapidly due to the frictional forces acting on NS1 as it moves through the envelope of the companion. The lost orbital energy is deposited into the gas envelope, causing it to become gravitationally unbound and get ejected, leaving behind the companion's bare helium core. The helium star eventually explodes in a supernova and becomes NS2, orbiting the recycled, milli-second pulsar NS1 in a compact binary orbit. Mass transfer from the helium star to NS1 can also occur prior to the second supernova explosion.

Here, our goal is to use the current DNS binary properties listed in Table 1 to constrain as best possible the magnitude of the natal kick and the mass of the progenitor of NS2 in the 8 known DNS in our galaxy at the time of the second supernova. We also predict the unknown radial velocity of each system, as well as the total transverse velocity and NS1 spin tilt angle for those systems where these properties have not been measured yet. We achieve these goals by means of Monte Carlo simulations, incorporating orbital dynamics during supernova explosions and kinematic of binaries in the Galactic potential, as detailed in what follows.

We start our analysis by deriving a probability distribution function (PDF) for the natal kick of NS2, which depends on the semi-major axis and orbital eccentricity immediately after the second supernova ($A_{postSN}$ and $e_{posSN}$ respectively). As the tidal interaction between two neutron stars in DNS is extremely weak (Bildsten \& Cutler 1992), the major mechanism that changes the orbital semi-major axis and eccentricity of DNS is the emission of gravitational waves. To account for gravitational radiation driven orbital evolution, we use the current observational parameters as initial conditions and integrate equations (35) and (36) in Junker \& Sch\"{a}fer  (1992) backwards in time. Because of the uncertainty in the true age of the observed DNS (see Kiziltan et al. 2009 for details), we randomly draw the integration time $t_i$ from a uniform distribution of ages between 0 and $\tau_c$, where $\tau_c$ is the characteristic age of the DNS. If $\tau_c$ is greater than the age of our galaxy, which is about 10 Gyr , we use 10 Gyr as the upper limit for $t_i$ instead. For PSR B1534+12, PSR B1913+16, and PSR J0737-3039, different upper limits are adopted for $t_i$, as discussed in more detail in the next section.

Once $A_{postSN}$ and $e_{postSN}$ are known from the gravitational radiation integration, we randomly draw the natal kick velocity ($V_k$) from a uniform distribution of values between 0 and 2500 km/s (i.e., we assume no prior kick distribution derived from pulsar samples analysis), and the kick direction $(\theta, \phi)$ from an isotropic distribution. Here $\theta$ is the polar angle between the natal kick velocity and the instantaneous orbital velocity of NS2's progenitor at the time of the second supernova explosion, and $\phi$ is the corresponding azimuthal angle (see Figure 1 in Kalogera 2000 for a graphical representation). Using the conservation laws of orbital energy and angular momentum expressed by equations (19) and (20) in Willems et al.  (2005), we obtain the orbital semi-major axis ($A_{preSN}$) immediately before the second supernova and the mass of NS2's helium star ($M_{2i}$). The pre-supernova orbit is assumed to be tidally circularized at the time of the second supernova explosion. On the other hand, the spin tilt angle of NS1, which is the angle between NS1 axis and the post supernova orbital angular momentum axis, can be calculated by following equation (3) in Kalogera (2000). Also, using equations (12) in the same paper, we find the polar angle $\theta'$ between the natal kick and the pre-supernova orbital angular momental axis. 

Immediately before the second supernova explosion, there may be mass transfer from NS2's progenitor to NS1. If the radius $R_{2i}$ of NS2's progenitor is greater than its Roche lobe radius $R_L$, which is approximated by equation (2) in Eggleton (1983), Roche lobe overflow (RLO) takes place.  Since NS2's progenitor is a bare helium star before the second supernova, we use equation (A10) in Kalogera \& Webbink (1998) to calculate $R_{2i}$. Note that the left hand side of that equation should be $\log R_{He,f}$ instead of $R_{He,f}$. In order to have dynamically stable RLO, so that NS2's progenitor does not coalesce with NS1 to form an isolated black hole, the mass ratio $M_{2i}$ to $M_1$ needs to be less than 3.5 (Ivanova et al. 2003). Even if NS2's progenitor does not overflow its Roche lobe, there may be mass transfer by stellar wind. In addition, $M_{2i}$ is restricted to be less than 8 M$_{\sun}$, which is a conventional limit that a helium star will become a neutron star instead of a black hole (see, e.g., Figure 1 in Belczynski et al. 2002; Table 16.4 in Tauris \& van den Heuvel 2004). 

The conservation laws of orbital energy and angular momentum allow for real solutions only for $M_{2i}$, $V_k$, $\theta$, $\phi$, and $A_{preSN}$ values satisfying the orbital dynamics constraints given by equation (21)--(27) in Willems et al. (2005). For PSR B1534+12, PSR B1913+16, and PSR J0737-3039, these constraints are supplemented with the measured NS1 spin tilt angle constraints. Applying these constraints together with the constraints imposing stability of any mass transfer taking place right before the second supernova explosion yields a PDF of the natal kick velocity of NS2. A corresponding PDF for the kick velocity $V_{k,sys}$ imparted to the binary's center of mass is obtained using equations (28)--(32) in Willems et al. (2005). In all past studies, kinematic analyses have either been skipped (Wang et al. 2006) or simplified to one dimension (Piran \& Shaviv 2004) or have suffered by ad hoc assumptions about the un-measurable radial velocities. Willems et al. (2006) used general DNS population synthesis models to guide this assumption. In this sutdy we employ a new method that still uses forward in time kinematic simulations, but they are tied to each observed system instead of being general.

We perform a Monte Carlo simulation of the motion of DNS in the Galactic potential. For this purpose, we adopt a reference frame with origin at the Galactic center, with $z$-axis pointing to the northern Galactic pole, and with the $x$-axis pointing in the direction from the Sun to the Galactic center. With respect to this reference frame, the Sun is located at $\vec{R}_\odot = (-8.5, 0, 0.03)$ \,kpc (Ghez et al. 2008, Gillessen et al. 2009, Reid et al. 2009, Joshi 2007) and has a peculiar velocity of $(10, 5, 7)$\,km/s (Bienaym\'e 1999). With the given $\vec{R}_\sun$ and peculiar velocity, the Sun is moving at a velocity $\vec{V}_\odot = (10, 226, 7)$ \,km/s.

To simulate the motion of DNS in the Galaxy, we randomly distribute a population of newly formed DNS in the Galactic disk according to a double exponential distribution function
\begin{equation}
n(R,z) = n_0 \exp \left( \frac{-R}{h_R} \right ) \exp \left ( \frac{-|z|}{h_z} \right ) .
\end{equation}
Here R and z are the cylindrical galactic coordinates, $h_R$ and $h_z$ are the Galactic scale length and height respectively. The distribution is normalized to unity by setting $n_0 = 1/(4 \pi h_z h_R^2)$. Since the progenitors of DNS are distributed in the same way as the massive stellar binaries in our galaxy, we choose $h_R$ = 2.8 kpc and $h_z$ = 0.07 kpc (see Joshi 2007). For all of the observed DNS except PSR B1534+12, we use a population of $\sim$10$^9$ simulated binaries. We use a population of $\sim$10$^{12}$ in the analysis of PSR B1534+12, due to the limiting chance of kicking it to the currently high Galactic altitude and simultaneously satisfying the constraints derived from the orbital dynamics of supernova explosions. If the scale height $h_z$ were larger than 70 pc, it would reduce the difficulty in satisfying the galactic dynamics constraints of PSR B1534+12. A smaller population of simulated binaries would then be needed for the analysis of this system.

The initial center-of-mass velocity of the DNS is obtained by summing up the kick velocity $\vec{V}_{k,sys}$ imparted to the binary center of mass during the second supernova explosion and the local Galactic rotational velocity $\vec{V}_{rot}$. Given the unknown orientation of the binary orbit in the Galactic reference frame, the systemic kick velocity $\vec{V}_{k,sys}$ is assumed to be distributed isotropically in space. The magnitude of $\vec{V}_{k,sys}$ is drawn from the velocity distribution derived in the previous step specific to each of the observed systems.

Starting from the randomly generated initial position and velocity, the motion of the DNS is calculated by numerically integrating the equations of motion from the DNS birth time to the current epoch. The equations of motion are derived from the Galactic potential of Carlberg \& Innanen (1987) with updated model parameters of Kuijken \& Gilmore (1989). The DNS birth time is chosen according to the DNS age $t_i$ generated in the first step of the analysis.

Lastly, we use the measured current position of the observed DNS to further constraint the ranges of possible $V_k$ and $M_{2i}$. We set the tolerance of this constraint to be within 100 pc from the observed position. We test the dependence of our results on this tolerance level by repeating the analysis with a 10 pc, or 50 pc tolerance, but our results do not show any significant changes. For PSR J1518+4904 and PSR J1756-2251, the position constraint is supplemented with the measured current proper motion constraint. For PSR B1534+12, PSR B1913+16, and PSR J0737-3039, the position constraint is supplemented with both the measured proper motion and NS1 spin tilt angle constraints.

\section{Results}

The analysis outlined in the previous section allows us to derive PDFs for the kick velocity $V_k$, NS2's immediate pre-supernova progenitor mass $M_{2i}$, the spin tilt $\theta_t$ of NS1, the systemic radial velocity $V_{rad}$, and the current transverse velocity $V_{trans}$. Our primary interest is in $V_k$ and $M_{2i}$, because they can shed light on the type of supernova that formed NS2, which will be discussed in more detail in the next section. The remaining quantities give us some predictions on currently unavailable parameters of the 8 known DNS in our galaxy. Instead of focusing on the most likely values as studies have done in the past, we derive confidence levels for the unknown parameters and they are summarized in Table 2.

Since the $V_k$ and $M_{2i}$ constraints are correlated, we present our results in the form of confidence level plots of a 2 dimensional joint probability distribution of $V_k$ - $M_{2i}$, for each DNS binary, in addition to the corresponding 1 dimensional PDF's of  $V_k$ and $M_{He}$. 
As pointed out by Willems et al (2006), physical parameter constraints are affected by the dimensionality of the PDF used, because of inherent correlations between parameters and projection effects. This variation is stronger when one solely examines the most likely values (i.e., values at the peak of the PDF) and weakens as broader confidence levels are considered. In what follows, the confidence levels of $V_k$ and $M_{2i}$ are derived from the 2D joint PDF. To calculate the 2D confidence levels, we first bin the data into a 2D grid and normalize the total probability within the 2D grid to unity. Then, we add the probability of each bin from the highest to the lowest, until the sum best matches the desired confidence level. On the other hand, the ranges of $V_{rad}$, $V_{trans}$, and $\theta_t$ are derived from the corresponding 1D PDF. In this case, the confidence level is found by having the shortest range of bins that has a total probability that best matches the desired confidence level.

\begin{deluxetable*}{lccccc}
\tablewidth{18.0 cm}
\tabletypesize{\small}
\tablecolumns{6}
\tablecaption{Possible Birth Properties and Current Kinematics Parameters of the 8 known DNS in our galaxy}
\tablehead{ 
\colhead{System} & \colhead{$V_k$ (km s$^{-1}$)} & \colhead{$M_{2i}$ ($M_{\sun}$)\tablenotemark{a}} & \colhead{$V_{rad}$ (km s$^{-1}$)\tablenotemark{b}} & \colhead{$V_{trans}$ (km s$^{-1}$)\tablenotemark{c}} & \colhead{$\theta_t$ (deg)\tablenotemark{d}} }

\startdata

PSR B1534+12 & 170 - 260 (150 - 270) & 2.00 - 2.90 (1.34 - 3.40) & $-55$ - 50 ($-110$ - 145) & 121 $\pm$ 6 \tablenotemark{e} & 25.0 $\pm$ 3.8 \tablenotemark{e}\\
PSR B1913+16 &  200 - 410 (190 - 450)  & 1.40 - 3.30 (1.39 - 5.00) & $-95$ - 85 ($-255$ - 315) & 135 $\pm$ 25 \tablenotemark{e} & 18 or 162 $\pm$ 6 \tablenotemark{e}\\
PSR J0737-3039 & 5 - 50 (5 - 120) & 1.25 - 1.55 (1.25 - 1.90)  & 8 - 44 ($-20$ - 76) & $23.8^{+8.8}_{-6.4} \tablenotemark{e}$ & 0 - 2.0 ( 0 - 7.8 )\\
PSR J1518+4904 & 20 - 80 (5 - 110) & 1.80 - 3.30 (1.49 - 4.70) & $-30$ - 8 ($-62$ - 48) & 25.4 $\pm$ 3.6 \tablenotemark{e} & 2 - 18 (0 - 32)\\ 
PSR J1756-2251 & 5 - 80 (5 - 185) & 1.25 - 1.90 (1.25 - 2.65) & $-34$ - 32 ($-110$ - 98) & 6.5 - 28.5 (6.0 - 64.5) & 0 - 3.4 ( 0 - 16.4 ) \\ 
PSR J1811-1736 & 0 - 170 (0 - 310) & 1.11 - 4.00 (1.11 - 8.00) & 0 - 112 ($-90$ - 210) &  30 - 114 (4 - 202) & 0 - 10.2 ( 0 - 57.6 ) \\
PSR J1829+2456 & 5 - 85 (5 - 225) & 1.40 - 2.70 (1.36 -  6.10) & $-30$ - 26 ($-126$ - 66) & 12 - 54 (4 - 102) & 0 - 8.5 ( 0 - 30.0 ) \\
PSR J1906+0746 & 5 - 170 (5 - 510) &  1.25 - 2.90 (1.25 - 4.80) & 12 - 103 ($-176$ - 297) & 89 - 174 ( 21 - 376) & 0 - 6.7 ( 0 - 51.5 ) \\

\enddata

\tablenotetext{a}{NS2's immediate pre-supernova progenitor mass}
\tablenotetext{b}{Systemic radial velocity}
\tablenotetext{c}{Current transverse velocity}
\tablenotetext{d}{Spin tilt angle of NS1}
\tablenotetext{e}{quantities obtained or derived from observations}
\tablecomments{The numbers without parenthesis in each entry are at 60\% confidence, whereas those with parenthesis are at 95\% confidence}

\end{deluxetable*}

\subsection{PSR B1534+12}

In this binary, the masses of NS1 and NS2 are 1.33 $M_{\sun}$ and 1.35 $M_{\sun}$ respectively. The orbital period is $P_b = 0.421$ days, with an eccentricity $e = 0.274$. The angle between NS1's spin axis and the orbital angular momentum axis is $25.0^\circ \pm 3.8^\circ$ or $155.0^\circ \pm 3.8^\circ$. The binary is currently located out of the galactic plane, at $l = 19.8^\circ$ and $b = 48.3^\circ$, and 1.02 kpc away from us. The observed proper motion $\mu_{R.A.} =  1.3$ mas yr$^{-1}$ and $\mu_{dec.} =  25.2$ mas yr$^{-1}$. At the measured distance this implies $V_{R.A.} = 6.47 \pm 0.32$ km s$^{-1}$ and $V_{dec.} = -121 \pm 6$ km s$^{-1}$, in the reference frame of the Sun. Also, we use the spin down age, which is 210 Myr given by Arzoumanian et al. 1999, as the upper limit of PSR B1534+12 's age, instead of the characteristic age.

Furthermore, the current orientation of the orbital angular momentum axis is also constrained for this sytem. The orientation axis can be described by the inclination angle $i$ and the angle of the orbital ascending node on the plane of the sky ($\Omega$). The sine of the inclination angle is $\sin i$ = 0.975 (Stairs et al. 2004). The angle $\Omega$ is $70^\circ \pm 20^\circ$ or $290^\circ \pm 20^\circ$, reckoned north through east. The two solutions correspond to $\cos i < 0$ and $\cos i > 0$ respectively (Bogdanov et al. 2002). Since tidal effects between the two neutron stars are negligible, the orbital angular momentum axis keeps a fixed orientation in space. From the kinematic analysis of the Galactic motion, the constraints on the present-day orbital inclination and proper motion can therefore be used to determine the systemic kick component $V_{k,sys}^\parallel$ parallel to the post-supernova orbital angular momentum axis. Following Kalogera (1996) and Wex et al. (2000), 
$V_{k,sys}^{\parallel}$ can also be expressed analytically as

\begin{equation}
\frac{V_{k,sys}^{\parallel}}{V_r} = \sqrt{\kappa_1} \sin \theta_t \, = \frac{v_{kz}\sqrt{\kappa_1}}{\sqrt{{v_{kz}^2 + (v_{ky}+1})^2}}\, ,
\end{equation}
where
\begin{equation}
V_r = \sqrt{ \frac {G(M_{2i}+M_1)}{A_{preSN}}} \, ,
\end{equation}
\begin{equation}
v_{kj} = \frac{V_{kj}}{V_r}\, ,
\end{equation}
\begin{equation}
\kappa_1 = \frac{M_{2i}^2}{(M_{2i} + M_1)^2}\, .
\end{equation}
Here, $v_{kj}$ ($j = x, y, z$) is the $x$, $y$, or $z$ component of the natal kick velocity in the frame centered on NS2's progenitor (see figure 1 in Kalogera for a graphical representation). Note that the sign of $V_{k,sys}^{\parallel}$ needs to be consistent with the sign of $v_{kz}$. Hence, through equation (2), the $V_{k,sys}^{\parallel}$ derived from observations gives us a constraint on the y and z components of the natal kick, as well as $\theta_t$.

\begin{figure}
\begin{center}
\resizebox{7.0cm}{13.5cm}{\includegraphics{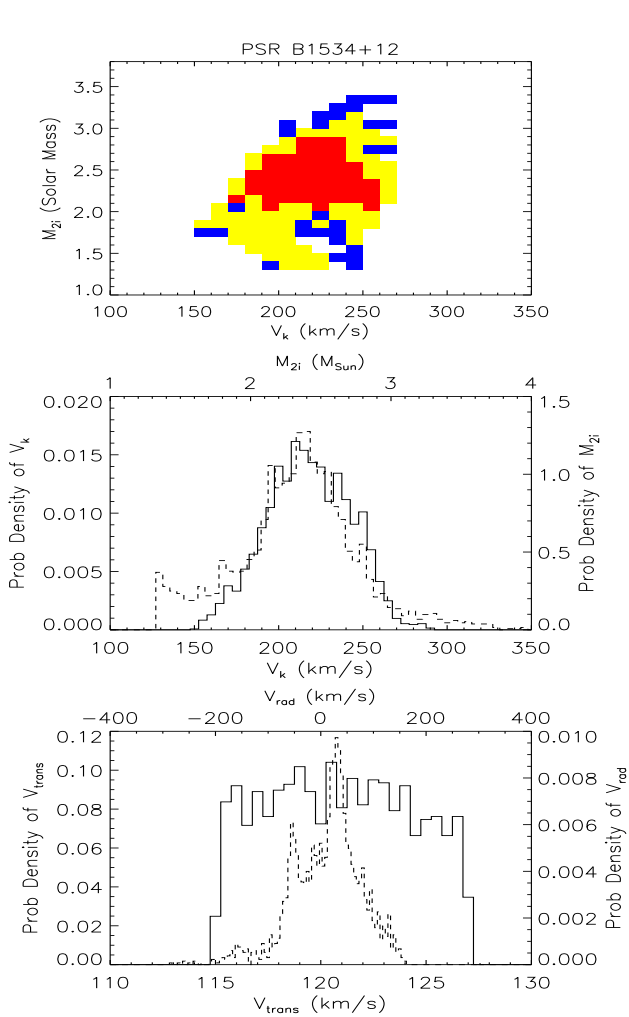}}
\end{center}
\caption{Kick velocity and progenitor mass of NS2, and present-day kinematic properties of PSR B1534+12.
$top$: The 2D joint $V_k$-$M_{2i}$ probability distribution. The red, yellow and blue colors represent 60\%, 90\% and 95\% confidence levels, respectively. $middle$: The $V_k$ (solid line) and $M_{2i}$ (dashed line)  PDF. The x-axes of  the $V_k$ and $M_{2i}$ PDF are printed on the bottom and the top of the plot respectively. $bottom$: The $V_{trans}$ (solid line) and $V_{rad}$ (dashed line) PDF. The x-axis of  the $V_{trans}$ is printed on the bottom, and x-axis of $V_{rad}$ PDF is on the top of the plot.}
\end{figure}

\begin{figure}
\begin{center}
\resizebox{7.0cm}{6.0cm}{\includegraphics{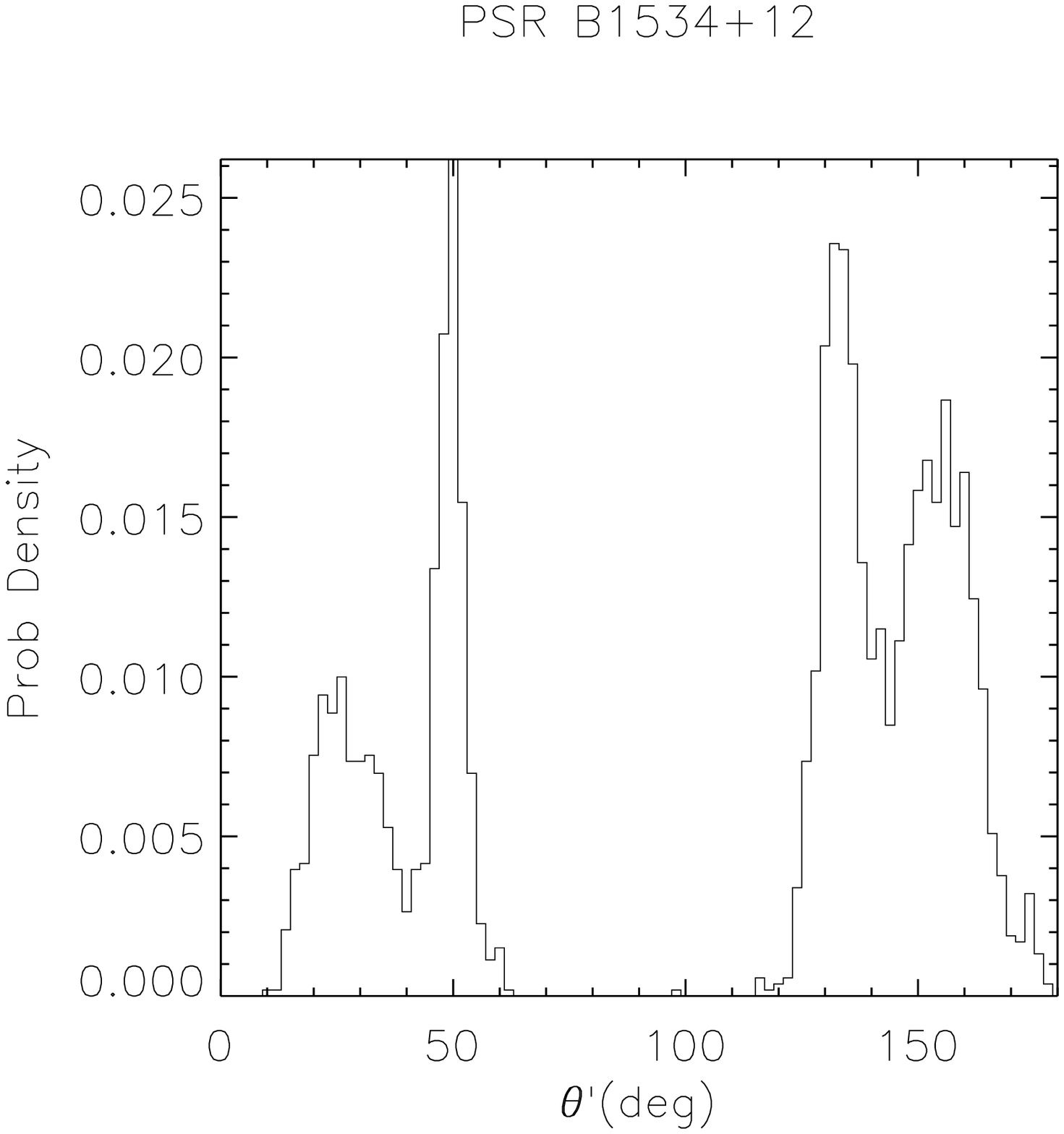}}
\end{center}
\caption{The  $\theta'$ PDF 's of PSR B1534+12.}
\end{figure}

The results are shown in Figures 1 and 2. Right before the second supernova, there must have been RLO from the progenitor of NS2 to NS1, as noted previously by Thorsett et al. (2005). As seen in the 2D $V_k$-$M_{2i}$ joint probability distribution in figure 1, the kick imparted to NS2 at birth is between 150 and 270 km/s, while the progenitor mass $M_{2i}$ is between 1.34 and 3.40 M$_\sun$, both at 95\% confidence. As seen in figure 2, polar kicks (i.e., $\theta' = 0^\circ$ or $180^\circ$) are unlikely, and the possible kick directions are asymmetric about the pre-supernova orbital plane (i.e., $\theta'$ = 90$^\circ$).  This asymmetry comes from the constraint on the orientation of the post-supernova orbital angular momentum axis on the sky. Moreover, we arrive at the same conclusion as Thorsett et al. (2005) that $\theta_t \sim 155^\circ$ is very unlikely. The current radial velocity of this binary is between -110 and 145 km/s, at 95\% confidence, and it is likely to be moving away from us. Within the observationally constrained range, the $V_{trans}$ PDF is roughly flat.

When comparing with recent analyses by Willems et al. (2004), Thorsett et al. (2005), and Wang et. al. (2006), our limits on $V_k$ and $M_{2i}$ are consistent wtih Thorsett et al. (2005), and are more constrained than Willlems et al. (2004) and Wang et al. (2006), because Willems et al. (2004) did not have the spin tilt $\theta_t$ measurement and Wang et al. (2006) did not have the proper motion and position constraints in their analysis. Our limits on the angle $\theta'$, which is the polar angle between the natal kick direction and the pre-supernova orbital angular momentum, are not as constrained as the limits shown in Thorsett et al. (2005). Through recent further investigation, it has been found that this difference is due to a code typo related to the galactic latitude of the source, which fortunately affected significantly only the $\theta'$ results (Dewey \& Stairs 2010, private communication).

\subsection{PSR B1913+16}

The masses of NS1 and NS2 in this binary are 1.44 $M_{\sun}$ and 1.38 $M_{\sun}$ respectively. The orbital period is $P_b = 0.323$ days, with an eccentricity $e = 0.617$. The angle between the NS1 spin axis and the post supernova orbital angular momentum axis is $18^\circ \pm 6^\circ$ or $162^\circ \pm 6^\circ$. At the current time, this binary is about 300 pc away the galactic plane, at $l = 50.0^\circ$ and $b = 2.1^\circ$, and 8.3 kpc away from us. The observed proper motion in right ascension and declination is $\mu_{R.A.} =  -3.27$ mas yr$^{-1}$ and $\mu_{dec.} =  -1.04$ mas yr$^{-1}$. At the measured distance, this implies $V_{R.A.} = -129 \pm 25$ km s$^{-1}$ and $V_{dec.} = -41 \pm 18$ km s$^{-1}$, in the reference frame of the Sun. In addition, instead of the characteristic age, we use the spin down age, which is 80 Myr (Arzoumanian et al. 1999),  as the upper limit on PSR B1913+16 's age.

\begin{figure}
\begin{center}
\resizebox{7.0cm}{13.5cm}{\includegraphics{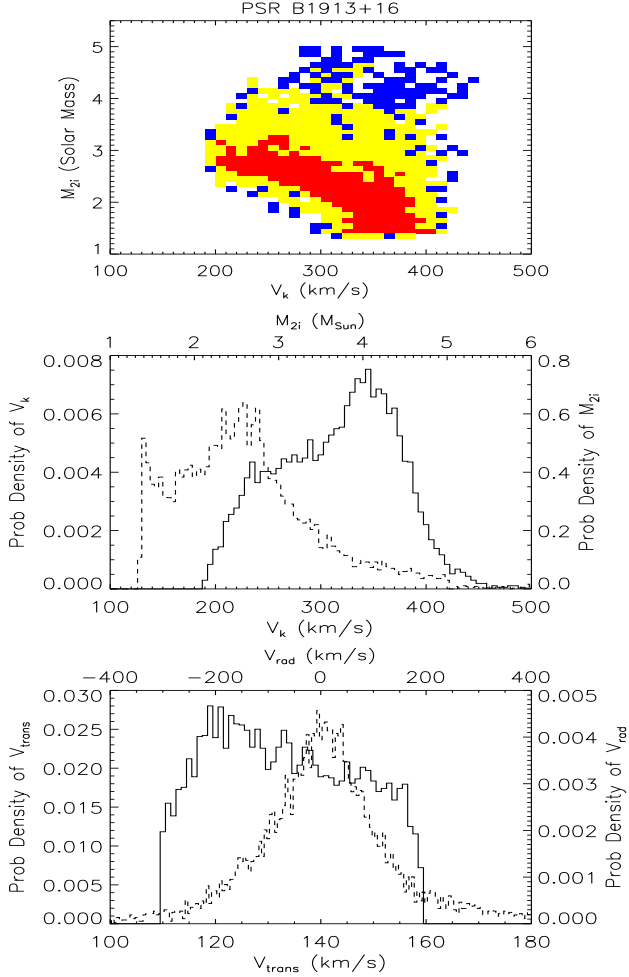}}
\end{center}
\caption{Kick velocity and progenitor mass of NS2, and present-day kinematic properties of PSR B1913+16.
$top$: The 2D joint $V_k$-$M_{2i}$ probability distribution. The red, yellow and blue colors represent 60\%, 90\% and 95\% confidence levels, respectively. $middle$: The $V_k$ (solid line) and $M_{2i}$ (dashed line)  PDF. The x-axes of  the $V_k$ and $M_{2i}$ PDF are printed on the bottom and the top of the plot respectively. $bottom$: The $V_{trans}$ (solid line) and $V_{rad}$ (dashed line) PDF. The x-axis of  the $V_{trans}$ is printed on the bottom, and x-axis of $V_{rad}$ PDF is on the top of the plot.}
\end{figure}

\begin{figure}
\begin{center}
\resizebox{7.0cm}{6.0cm}{\includegraphics{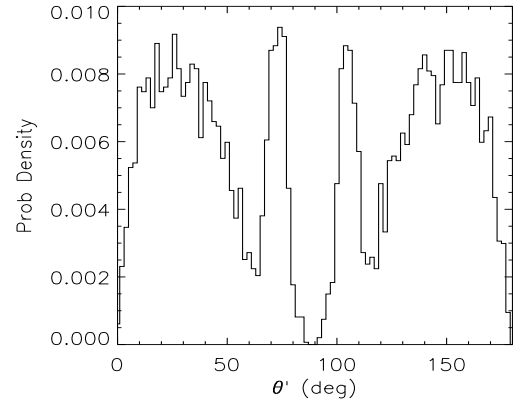}}
\end{center}
\caption{The  $\theta'$ PDF 's of PSR B1913+16.}
\end{figure}

The results are displayed in Figures 3 and 4. As shown in the 2D $V_k$-$M_{2i}$ joint probability distribution in figure 3, the kick imparted to NS2 at birth is between 190 and 450 km s$^{-1}$, while the progenitor mass $M_{2i}$ is between 1.4 and 5.0 M$_\sun$, both at 95\% confidence. As seen from figure 4, both polar and planar kick directions are allowed. The current radial velocity is between $-255$ and 315 km s$^{-1}$ at 95\% confidence. Within the observationally constrained range, the $V_{trans}$ PDF is roughly flat, with a slight preference for transverse velocities of $\sim$120 km/s. Furthermore, although observations show a degeneracy in $\theta_t$ at $\sim\,18^\circ$ and $\sim\,162^\circ$, our analysis shows that $\theta_t$ is much more likely to be $\sim\,18^\circ$.

In addition, although our confidence intervals of $V_k$ and $M_{2i}$ are consistent with Willems et al. (2004) and Wang et. al. (2006), our reported intervals are tighter, because Willems et al. (2004) and Wang et al. (2006) reported the allowed ranges of $V_k$ and $M_{2i}$ instead of confidence intervals.

\subsection{PSR J0737-3039}

The masses of the two neutron stars in this binary are 1.34 and 1.25 M$_\sun$, where the more massive one is NS1. They are orbiting each other with a period of 0.102 days, and an eccentricity of 0.0878. At the current time, the binary is close to the galactic plane, at $l = 245.2^\circ$ and $b = -4.5^\circ$. Through direct measurement of geometric parallax, Deller et al. (2009) measured the distance to PSR J0737-3039 to be 1.15 kpc, and the proper motion in R.A. and declination to be $-$3.82 and 2.13 mas yr$^{-1}$ respectively. At the measured distance, this implies $V_{R.A.} = -20.8$ km s$^{-1}$ and $V_{dec.} = 11.6$ km s$^{-1}$, in the reference frame of the Sun. In addition, Lorimer et al. (2007) constrained the possible age range to be either $70 - 90$ or $170 - 190$ Myr. We adopt these age ranges in our analysis of PSR J0737-3039.

As the radio pulse profile of PSR J0737-3039A (i.e., NS1) has not changed significantly through six years of observation, the spin tilt angle of J0737-3039A with respect to the post supernova orbital angular momentum axis is believed to be small. However, there is no well constrained value for this angle. Ferdman et al. (2008) and Breton (2008) argue that there are two possible models to explain the observed pulse profile: the single-cone and two-cone emission model. In this context, the number of emission cones means the number of emitting cones that intersect with our line of sight. Each model gives a different spin tilt angle estimate. Ferdman et al. (2008) estimate this angle to be < 15$^\circ$ for the single-cone model and < 6.1$^\circ$ for the two-cone model, both at 68.3 \% confidence. Due to this uncertainty, we take the spin tilt angle $\theta_t$to be < 15$^\circ$ in our work, which corresponds to 68.3 \% confidence value in the single-cone model and 95.4 \% confidence value in the two-cone model in Ferdman et al (2008).

\begin{figure}
\begin{center}
\resizebox{7.0cm}{13.5cm}{\includegraphics{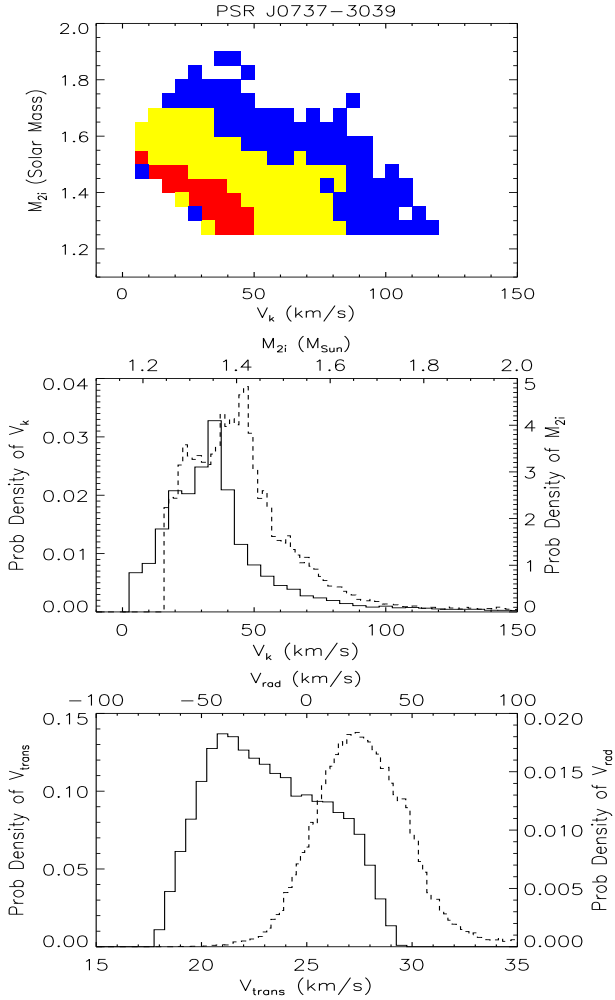}}
\end{center}
\caption{Kick velocity and progenitor mass of NS2, and present-day kinematic properties of PSR J0737-3039.
$top$: The 2D joint $V_k$-$M_{2i}$ probability distribution. The red, yellow and blue colors represent 60\%, 90\% and 95\% confidence levels, respectively. $middle$: The $V_k$ (solid line) and $M_{2i}$ (dashed line)  PDF. The x-axes of  the $V_k$ and $M_{2i}$ PDF are printed on the bottom and the top of the plot respectively. $bottom$: The $V_{trans}$ (solid line) and $V_{rad}$ (dashed line) PDF. The x-axis of  the $V_{trans}$ is printed on the bottom, and x-axis of $V_{rad}$ PDF is on the top of the plot.}
\end{figure}

\begin{figure}
\begin{center}
\resizebox{7.0cm}{6.0cm}{\includegraphics{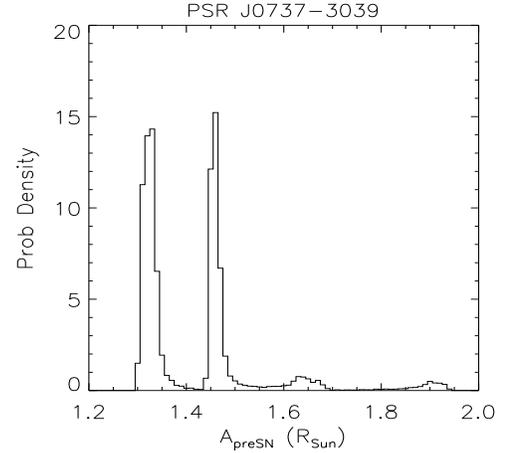}}
\end{center}
\caption{The orbital separation PDF of PSR J0737-3039 immediately before the second supernova.}
\end{figure}

\begin{figure}
\begin{center}
\resizebox{7.0cm}{6.0cm}{\includegraphics{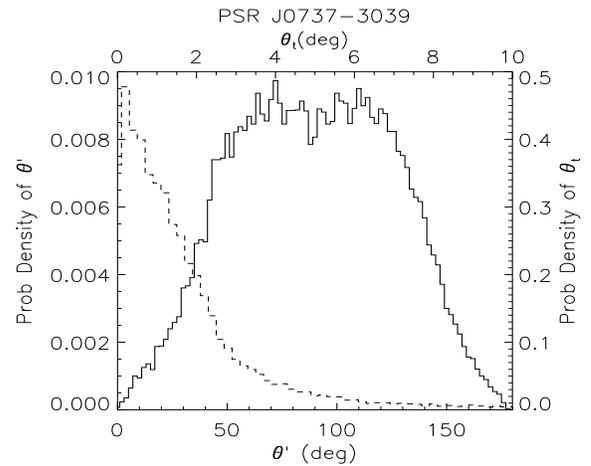}}
\end{center}
\caption{The  $\theta'$ (solid line) and $\theta_t$ (dash line) PDF 's of PSR J0737-3039. The x-axes of  the $\theta'$ and $\theta_t$ PDF are printed on the bottom and the top of the plot respectively.}
\end{figure}

The results are displayed in Figures 5-7. As shown in the 2D $V_k$-$M_{2i}$ joint probability distribution, the kick recieved by NS2 (i.e., PSR J0737-3039B) at birth is between 5 and 120 km s$^{-1}$, while the progenitor mass $M_{2i}$ is between 1.25 and 1.90 M$_\sun$, both at 95\% confidence. The probability distribution of the kick direction is symmetric about the pre-supernova orbital plane (i.e., $\theta^\prime$ = 90$^\circ$), and planar kicks are more favorable than polar kicks. The PDF of the semi-major axis immediately before the second supernova ($A_{preSN}$) peaks at two narrow ranges: $1.35 - 1.40$ and $1.45 - 1.50$ $R_\sun$. The first peak corresponds to an age of $70 - 90$ Myr, and the second peak to an age of $170 - 190$ Myr. At the current time, PSR J0737-3039 is more likely to be moving away from us, with a radial velocity between $-20$ and 76 km/s with 95\% confidence. Within the observationally constrained range, the $V_{trans}$ PDF peaks at $\sim$20 km/s. Furthermore, RLO from NS2's progenitor to NS1 before the second supernova explosion cannot be avoided.  

\begin{figure}
\begin{center}
\resizebox{7.0cm}{6.0cm}{\includegraphics{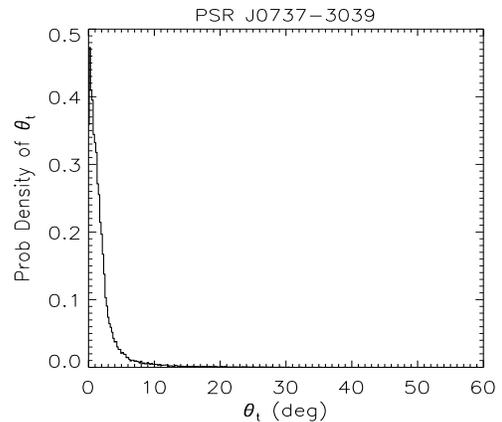}}
\end{center}
\caption{The  $\theta_t$ PDF  of PSR J0737-3039, obtained from the same constraints as before except for the upper limit of $\theta_t$, which is raised to 60$^\circ$, to test the sensitivity of our results  to the adopted $\theta_t$ upper limit.}
\end{figure}

To test the dependence of the presented results on the adopted upper limit for $\theta_t$, we repeated the analysis for the upper limits of $5^\circ$ and $60^\circ$. The results do not show any significant differences compared to those obtained with the $15^\circ$ upper limit. This is not surprising, because the $\theta_t$ PDF as shown in Figure 8, which is obtained from the same constraints as before except raising the upper limit of $\theta_t$ to 60$^\circ$, peaks strongly at 0$^\circ$ and decays exponentially with increasing values of $\theta_t$.

Comparatively, our confidence intervals on $V_k$ and $M_{2i}$ are consistent with Stairs et al. (2006) and Willems et al. (2006), but are much more constrained than Wang et al. (2006), mainly because they did not have the proper motion constraint in their analyses.   

\subsection{PSR J1518+4904}

The masses of the two neutron stars in this binary are 0.72 and 2.00 M$_\sun$, where the less massive one is NS1. However, each one has an uncertainty of  $\sim$0.5 M$_\sun$. In our analysis, we choose $M_1$ to be 1.23 and $M_2$ to be 1.49, so that $M_1$ is greater than the smallest measured neutron star mass, which is about 1.2 $M_\sun$. The characteristic age is 20 Gyr, which is larger than a Hubble time. Thus, we set the upper limit on PSR J1518's age to be 10 Gyr instead. The two neutron stars are orbiting each other with a period of 8.63 days, and an eccentricity of 0.249. At the current time, this binary is out of the galatic plane, at $l = 80.8^\circ$ and $b = 54.3^\circ$, and at a distance of 0.625 kpc from the Sun. The proper motion in R.A. and dec. is $\mu_{R.A.} = -0.67$ and  $\mu_{Dec.} = -8.53$ mas yr$^{-1}$. At the measured distance, this implies $V_{R.A.} = -1.99 \pm 0.31$ and $V_{Dec.} = -25.3 \pm 3.6$ km s$^{-1}$, in the reference frame of the Sun. There is no spin tilt measurement available for this system.

\begin{figure}
\begin{center}
\resizebox{7.0cm}{13.5cm}{\includegraphics{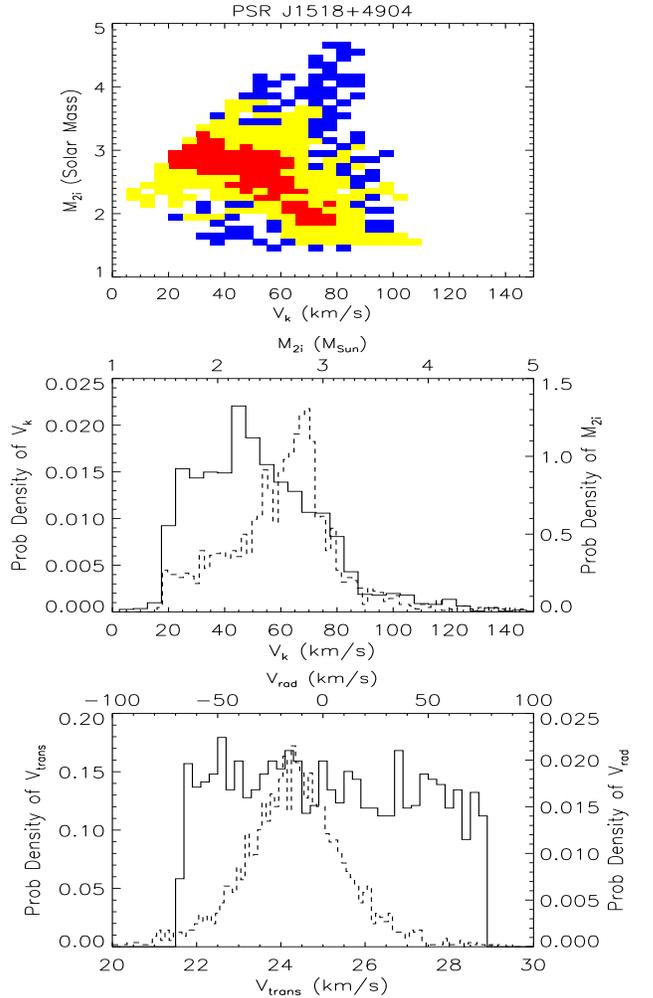}}
\end{center}
\caption{Kick velocity and progenitor mass of NS2, and present-day kinematic properties of PSR J1518+4904.
$top$: The 2D joint $V_k$-$M_{2i}$ probability distribution. The red, yellow and blue colors represent 60\%, 90\% and 95\% confidence levels, respectively. $middle$: The $V_k$ (solid line) and $M_{2i}$ (dashed line)  PDF. The x-axes of  the $V_k$ and $M_{2i}$ PDF are printed on the bottom and the top of the plot respectively. $bottom$: The $V_{trans}$ (solid line) and $V_{rad}$ (dashed line) PDF. The x-axis of  the $V_{trans}$ is printed on the bottom, and x-axis of $V_{rad}$ PDF is on the top of the plot.}
\end{figure}

\begin{figure}
\begin{center}
\resizebox{7.0cm}{6.0cm}{\includegraphics{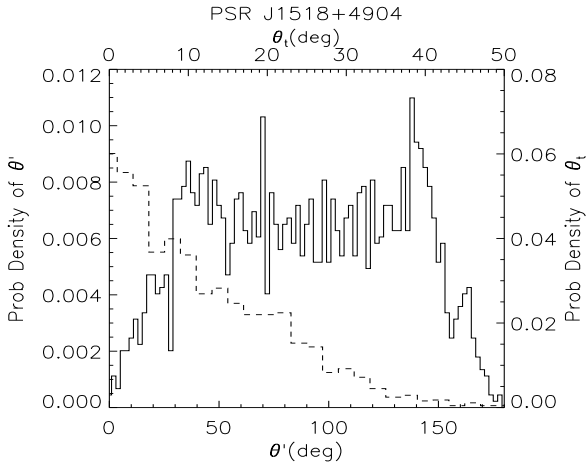}}
\end{center}
\caption{The  $\theta'$ (solid line) and $\theta_t$ (dash line) PDF 's of PSR J1518+4904. The x-axes of  the $\theta'$ and $\theta_t$ PDF are printed on the bottom and the top of the plot respectively.}
\end{figure}

The results are shown in Figures 9 and 10. As seen in the 2D $V_k$-$M_{2i}$ joint probability distribution, the kick imparted to NS2 at birth is between 5 and 110 km/s, while the progenitor mass $M_{2i}$ is between 1.5 and 4.7 M$_\sun$, both at 95\% confidence. Both polar and planar kick directions are possible, but the planar kicks have a higher probability. The spin tilt angle $\theta_t$ is smaller than 32$^\circ$ at 95\% confidence. The radial velocity is found to be between -62 and 48 km/s at 95\% confidence, and the system is currently likely moving towards us. Within the observationally constrained ranges, the $V_{trans}$ PDF is roughly flat. 

Our limits of $V_k$ and $M_{2i}$ are much more constrained than those of Wang et al. (2006), primarily because they did not have the proper motion and galactic position constraints included in their analysis.

\subsection{PSR J1756-2251}

The masses of the two neutron stars in this binary are 1.31 and 1.26 M$_\sun$, where the more massive one is NS1. The characteristic age is 443 Myr. The two neutron stars are in an orbit with a period of 0.320 days, and an eccentricity of 0.181. At the current time, this binary is close to the galactic plane, at $l = 6.5^\circ$ and $b = 0.95^\circ$, and at a distance of 2.5 kpc away from us.  Only proper motion $\mu_{R.A.}$ is measured, which is -0.7 mas/yr. At the measured distance, this implies $V_{R.A.} = -8.30 \pm 2.37$ km/s, in the reference frame of the Sun. There is currently no spin tilt measurement available for this system.

\begin{figure}
\begin{center}
\resizebox{7.0cm}{13.5cm}{\includegraphics{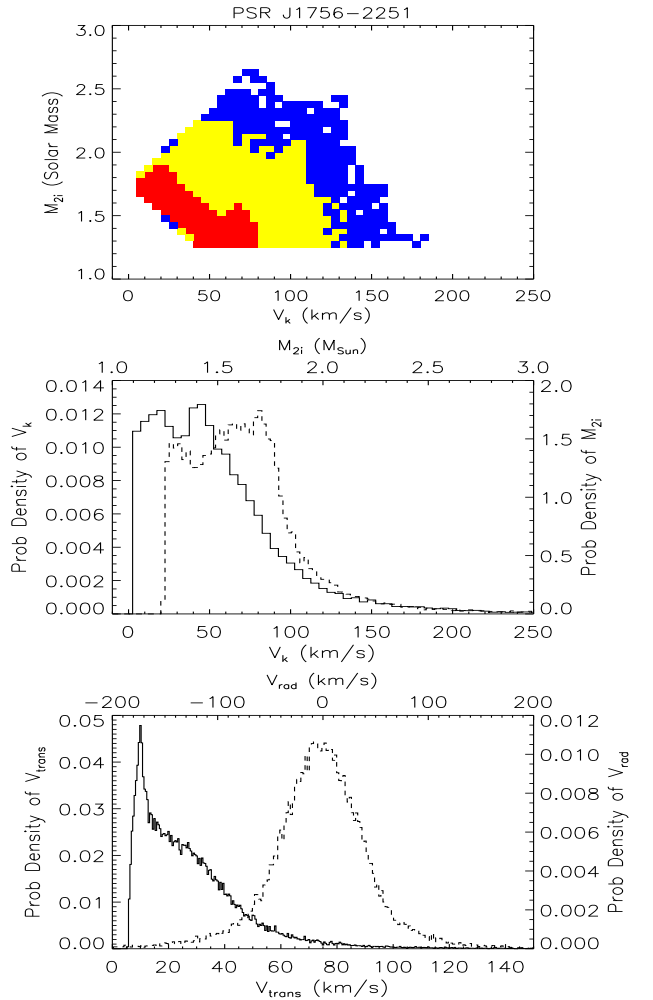}}
\end{center}
\caption{Kick velocity and progenitor mass of NS2, and present-day kinematic properties of PSR J1756-2251.
$top$: The 2D joint $V_k$-$M_{2i}$ probability distribution. The red, yellow and blue colors represent 60\%, 90\% and 95\% confidence levels, respectively. $middle$: The $V_k$ (solid line) and $M_{2i}$ (dashed line)  PDF. The x-axes of  the $V_k$ and $M_{2i}$ PDF are printed on the bottom and the top of the plot respectively. $bottom$: The $V_{trans}$ (solid line) and $V_{rad}$ (dashed line) PDF. The x-axis of  the $V_{trans}$ is printed on the bottom, and x-axis of $V_{rad}$ PDF is on the top of the plot.}
\end{figure}

\begin{figure}
\begin{center}
\resizebox{7.0cm}{6.0cm}{\includegraphics{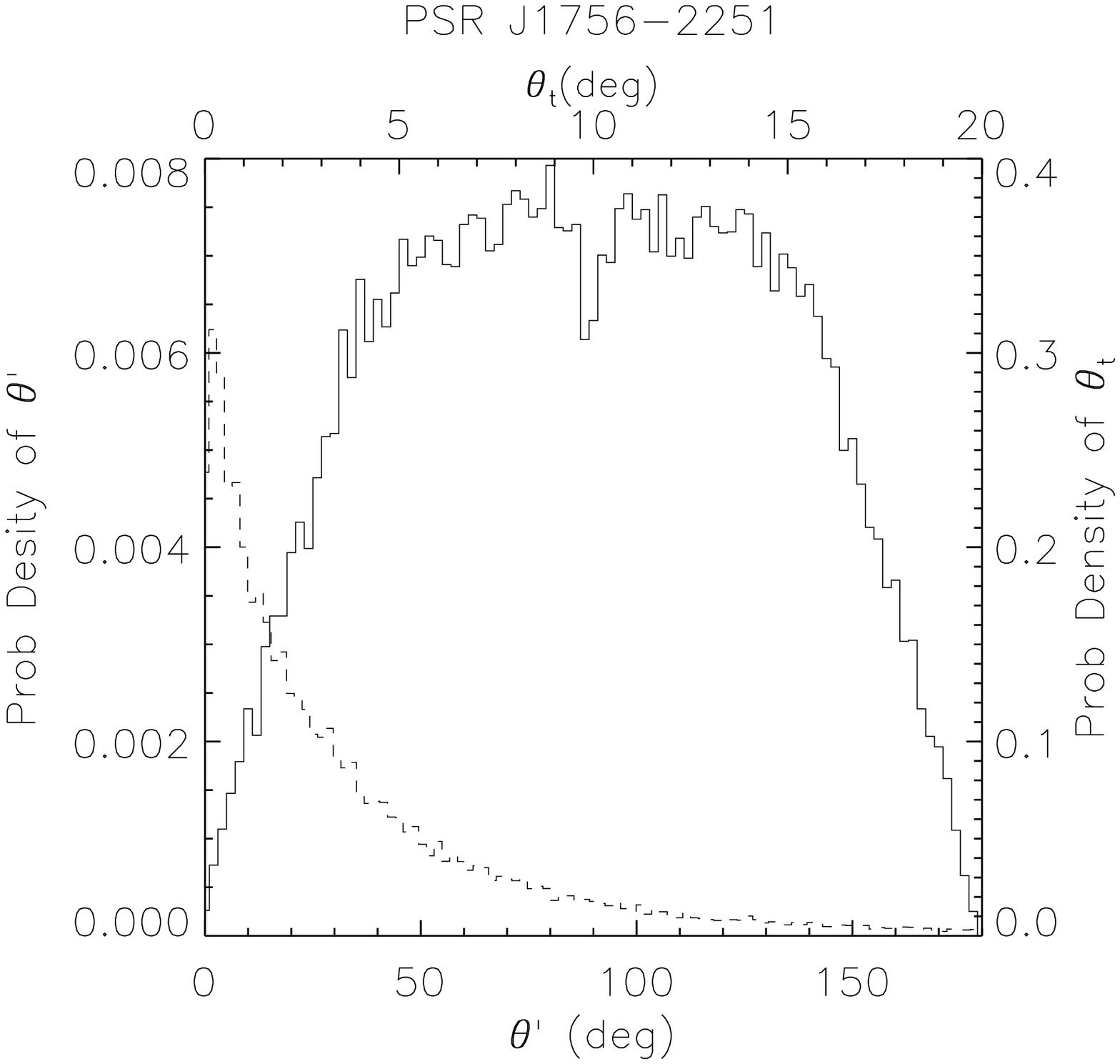}}
\end{center}
\caption{The  $\theta'$ (solid line) and $\theta_t$ (dash line) PDF 's of PSR J1756-2251. The x-axes of  the $\theta'$ and $\theta_t$ PDF are printed on the bottom and the top of the plot respectively.}
\end{figure}

The results are shown in Figures 11 and 12. From the 2D $V_k$-$M_{2i}$ joint probability distribution, the kick received by NS2 at birth is between 5 and 185 km s$^{-1}$, while the progenitor mass $M_{2i}$ is between 1.25 and 2.65 M$_\sun$, both with 95\% confidence. Both polar and planar kicks are allowed, but planar kicks are more favorable than polar ones. The spin tilt angle of NS1 is smaller than 16.4$^\circ$ with 95\% confidence. At the current time and at the measured distance, PSR J1756-2251 has a total transverse velocity between 6.0 and 64.5 km s$^{-1}$, and a radial velocity  between $-110$ and 98 km s$^{-1}$, both with 95\% confidence. Furthermore, this binary is equally likely to be moving towards us as away from us.

When comparing with Wang et al. (2006), we have tighter constrained limits on $V_k$ and $M_{2i}$, due to the fact that we have the additional constraint on the proper motion $\mu_{R.A.}$.

\subsection{PSR J1811-1736}

The two neutron stars have masses of 1.62 and 1.11 M$_\sun$, where the more massive one is NS1. However, each mass measurement has an uncertainty of $\sim$0.5 M$_\sun$. PSR J1811 has a characteristic age of 1830 Myr. The neutron stars are in an orbit with a period of 18.8 days, and an eccentricity of 0.828. At the current time, this binary is close to the galactic plane, at $l = 12.8^\circ$ and $b = 0.44^\circ$, and at a distance of 6.0 kpc away from us. Like PSR J1756-2251, there are currently no spin tilt nor proper motion measurements available. 

\begin{figure}
\begin{center}
\resizebox{7.0cm}{13.5cm}{\includegraphics{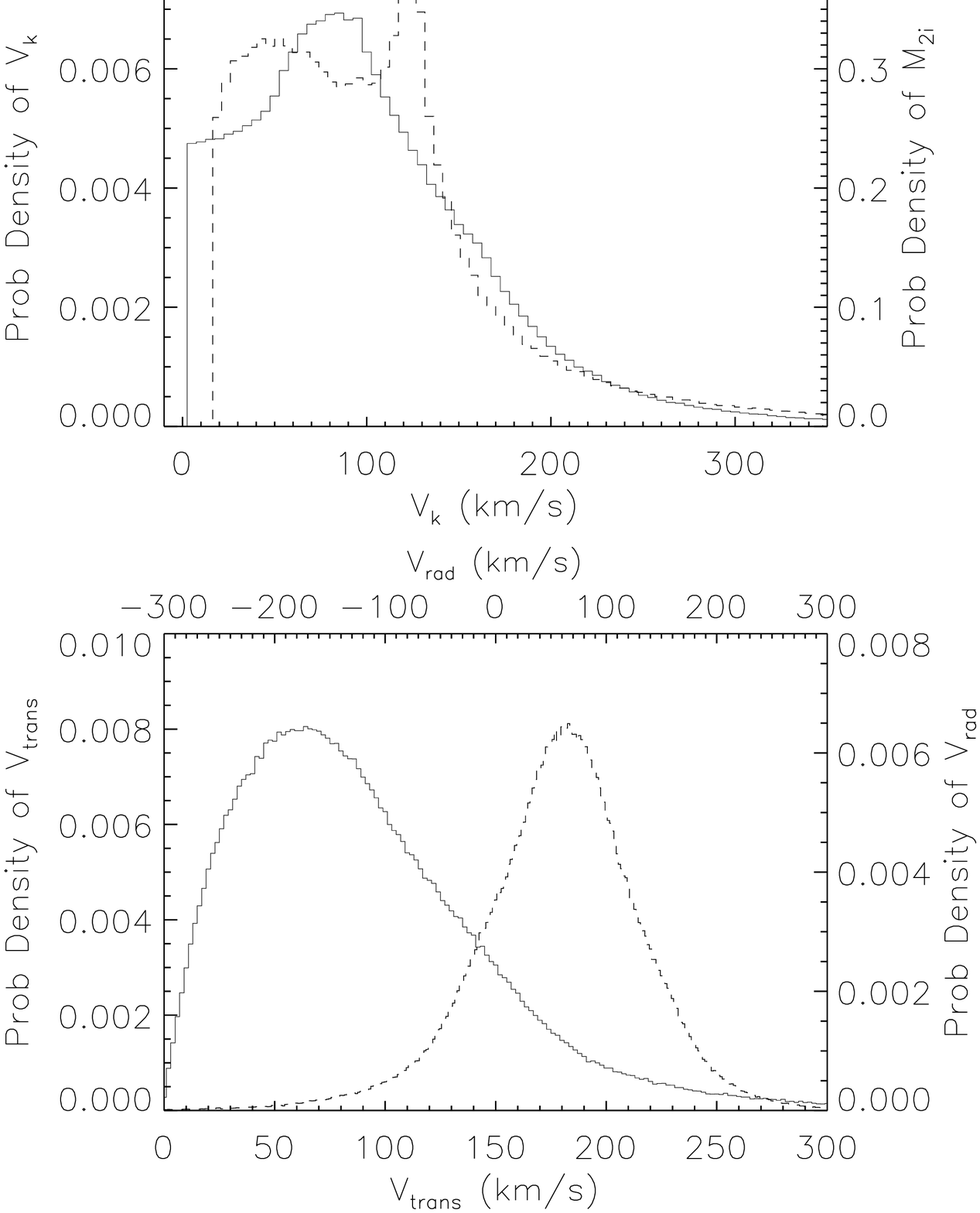}}
\end{center}
\caption{Kick velocity and progenitor mass of NS2, and present-day kinematic properties of PSR J1811-1736.
$top$: The 2D joint $V_k$-$M_{2i}$ probability distribution. The red, yellow and blue colors represent 60\%, 90\% and 95\% confidence levels, respectively. $middle$: The $V_k$ (solid line) and $M_{2i}$ (dashed line)  PDF. The x-axes of  the $V_k$ and $M_{2i}$ PDF are printed on the bottom and the top of the plot respectively. $bottom$: The $V_{trans}$ (solid line) and $V_{rad}$ (dashed line) PDF. The x-axis of  the $V_{trans}$ is printed on the bottom, and x-axis of $V_{rad}$ PDF is on the top of the plot.}
\end{figure}

\begin{figure}
\begin{center}
\resizebox{7.0cm}{6.0cm}{\includegraphics{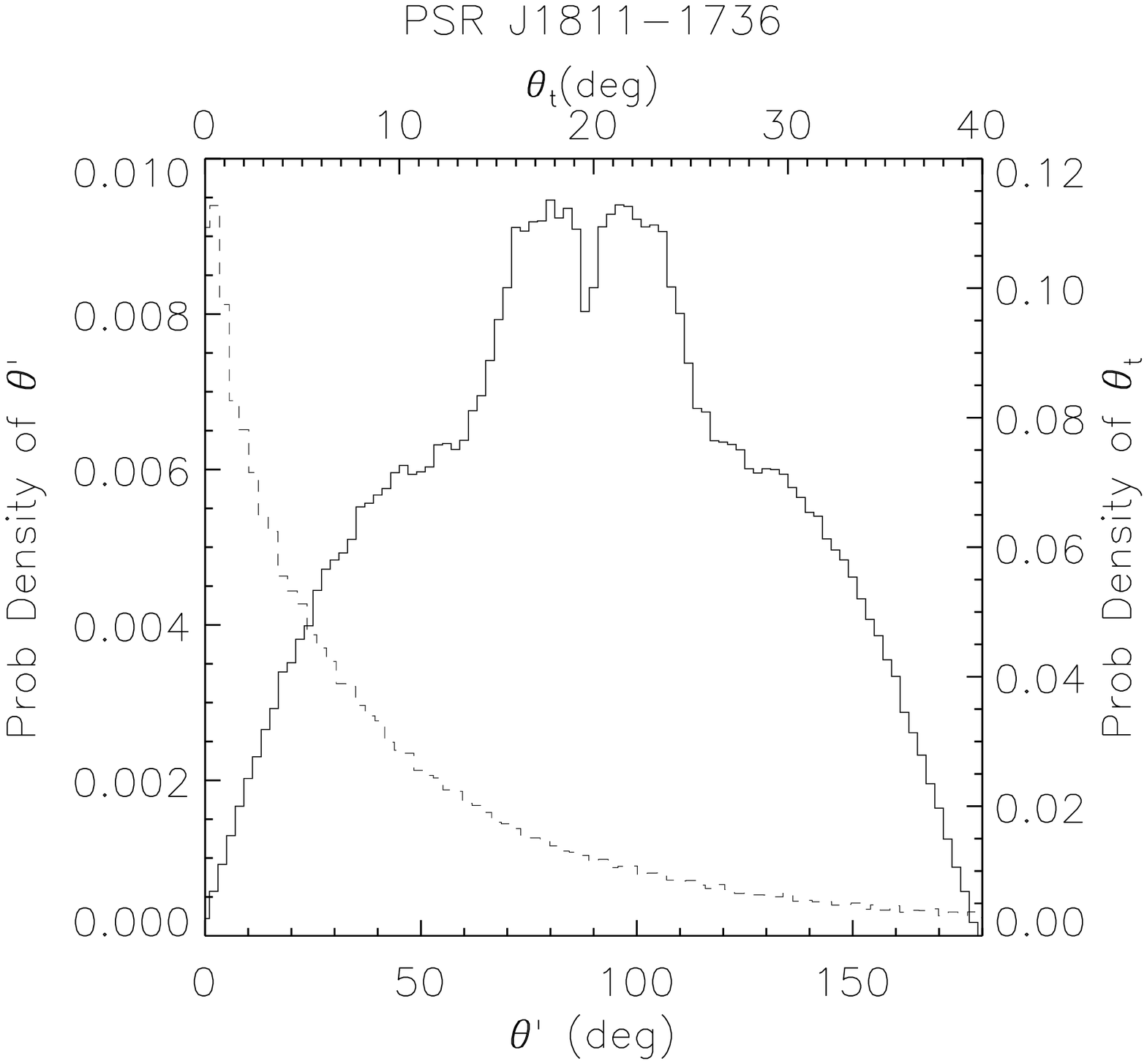}}
\end{center}
\caption{The  $\theta'$ (solid line) and $\theta_t$ (dash line) PDF 's of PSR J1811-1736. The x-axes of  the $\theta'$ and $\theta_t$ PDF are printed on the bottom and the top of the plot respectively.}
\end{figure}

The results are displayed in Figures 13 and 14. The 2D $V_k$-$M_{2i}$ joint probability distribution shows that the kick given to NS2 at birth is less than 310 km s$^{-1}$, while the progenitor mass $M_{2i}$ is between 1.11 and 8.0 M$_\sun$, both with 95\% confidence. Both polar and planar kicks are allowed, but the planar kicks are more favorable than the polar ones. The spin tilt angle $\theta_t$ is smaller than 57.6$^\circ$ with 95\% confidence. At the measured distance, the total transverse velocity of PSR J1811-1736 at the current epoch is between 4 and 202 km s$^{-1}$, and the radial velocity between -90 and 210 km s$^{-1}$ with 95\% confidence, which means this binary is likely moving away from us.

Although our upper boundary of the 95\% confidence intervals of $M_{2i}$ is the same as the upper limit found by Wang et al. (2006), the value of our lower boundary is less than their lower limit. This is because Wang et al. (2006) put a conservative lower limit of 2.1 M$_\sun$ on the progenitor mass of a neutron star, which we did not. On the other hand, our limits on $V_k$ are more constrained than those of Wang et al. (2006), as we compute confidence levels instead of allowed ranges of solution as Wang et al. (2006).

\subsection{PSR J1829+2456}

The two neutron stars of this binary have masses of 1.14 and 1.36 M$_\sun$, where the less massive one is NS1. However, there is an uncertainty of $\sim$0.5 M$_\sun$ in the mass measurements. The characteristic age is 12.4 Gyr. For the same reason as PSR J1518+4904, we set an the upper limit of 10 Gyr on the age of this binary in our analysis. The neutron stars are orbiting each other with a period of 1.18 days, and an orbital eccentricity of 0.139. At current time, this binary is out of the galactic plane, at $l = 53.3^\circ$ and $b = 15.6^\circ$, and it is 1.2 kpc away from us. Although there is no spin tilt nor proper motion measurements available at current time, Lorimer et al. (2005) derived that the transverse velocity is smaller than 118 km s$^{-1}$. 

\begin{figure}
\begin{center}
\resizebox{7.0cm}{13.5cm}{\includegraphics{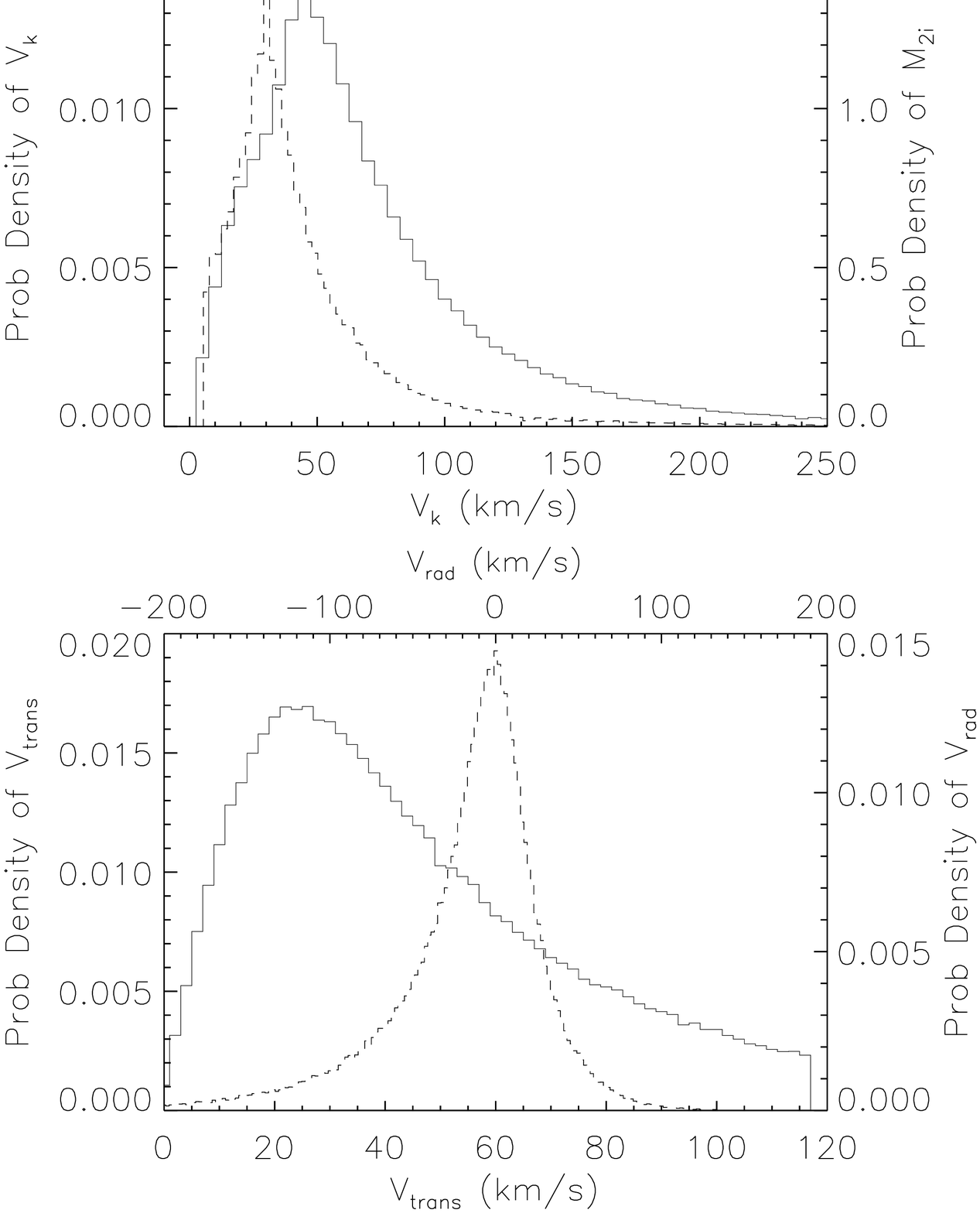}}
\end{center}
\caption{Kick velocity and progenitor mass of NS2, and present-day kinematic properties of PSR J1829+2456.
$top$: The 2D joint $V_k$-$M_{2i}$ probability distribution. The red, yellow and blue colors represent 60\%, 90\% and 95\% confidence levels, respectively. $middle$: The $V_k$ (solid line) and $M_{2i}$ (dashed line)  PDF. The x-axes of  the $V_k$ and $M_{2i}$ PDF are printed on the bottom and the top of the plot respectively. $bottom$: The $V_{trans}$ (solid line) and $V_{rad}$ (dashed line) PDF. The x-axis of  the $V_{trans}$ is printed on the bottom, and x-axis of $V_{rad}$ PDF is on the top of the plot.}
\end{figure}

The results are shown in Figure 15. The $V_k$-$M_{2i}$ joint probability distribution shows that the kick given to NS2 at birth is between 5 and 225 km s$^{-1}$, while the progenitor mass $M_{2i}$ is between 1.4 and 6.1 M$_\sun$ with 95\% confidence. The PDFs of $\theta_t$ and $\theta'$ are similar to those of PSR J1756-2251. Both polar and planar kicks are allowed, but planar kicks are more probable. The spin tilt angle $\theta_t$ is less than 30.0$^\circ$ with 95\% confidence. At the known distance, the total transverse velocity of PSR J1829+2456 at the current time is between 4 and 102 km s$^{-1}$, and the radial velocity between $-126$ and 66 km s$^{-1}$ with 95\% confidence. In addition, the probability that this system is moving towards the Sun is $\sim$20\% higher than that it is moving away from the Sun.

Comparing our $V_k$ and $M_{2i}$ results with Wang et al. (2006), we have more constrained limits, as they do not include any kinematic constraints.

\subsection{PSR J1906+0746}
The masses of the neutron stars are 1.37 and 1.25 M$_\sun$, where the more massive one is NS1. Unlike the other 7 binaries, the observed pulsar is NS2. This implies the binary is relatively young, with a charateristic age of 0.112 Myr only. The neutron stars are in an orbit with a period of 0.166 days, and an orbital eccentrirctiy of 0.0853. At current time, this binary is close to the galactic plane, at  $l = 41.6^\circ$ and $b = 0.15^\circ$, and it is 5.4 kpc away from us. Again, there are no spin tilt nor proper motion measurements available yet.

\begin{figure}
\begin{center}
\resizebox{7.0cm}{13.5cm}{\includegraphics{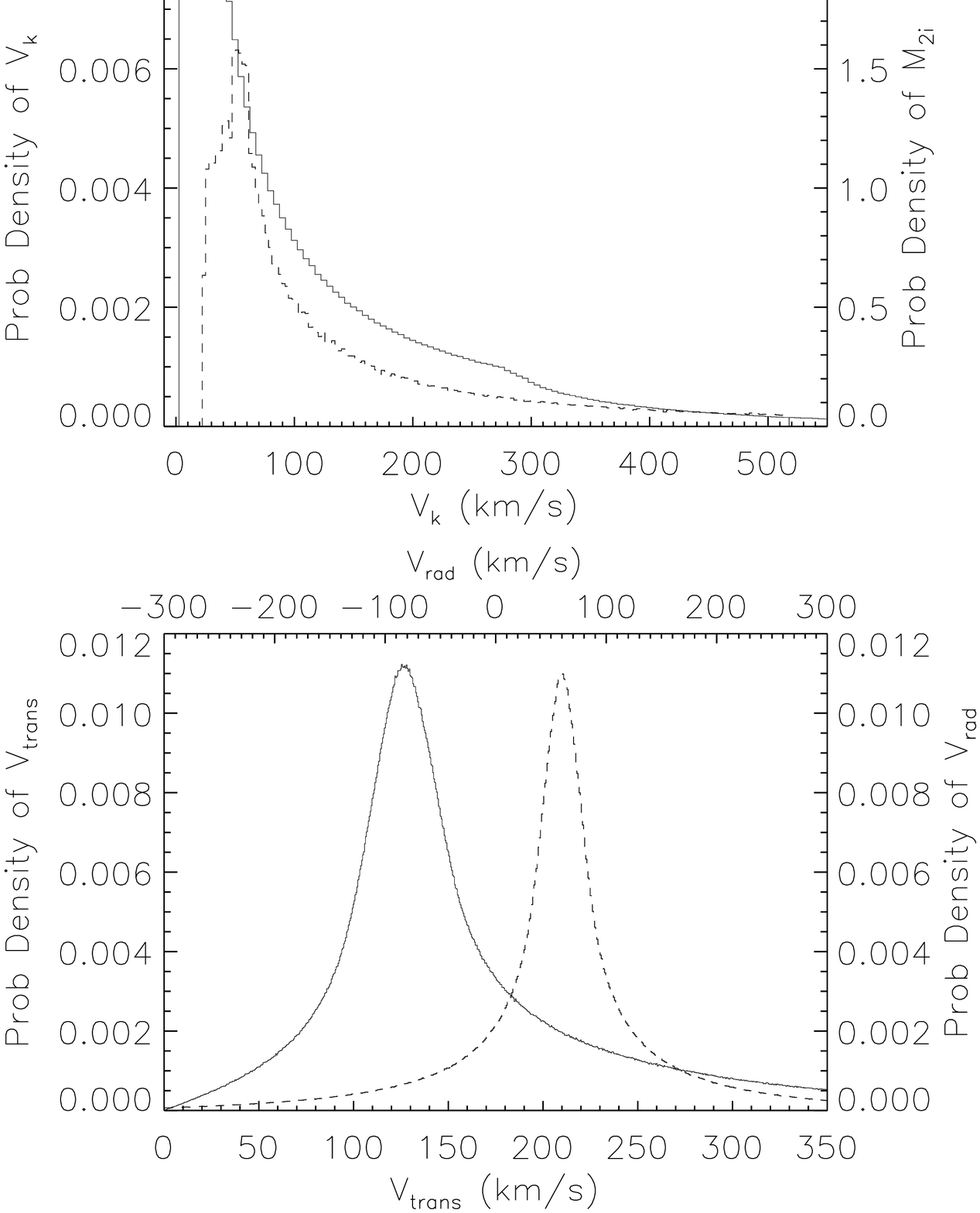}}
\end{center}
\caption{Kick velocity and progenitor mass of NS2, and present-day kinematic properties of PSR J1906+0746.
$top$: The 2D joint $V_k$-$M_{2i}$ probability distribution. The red, yellow and blue colors represent 60\%, 90\% and 95\% confidence levels, respectively. $middle$: The $V_k$ (solid line) and $M_{2i}$ (dashed line)  PDF. The x-axes of  the $V_k$ and $M_{2i}$ PDF are printed on the bottom and the top of the plot respectively. $bottom$: The $V_{trans}$ (solid line) and $V_{rad}$ (dashed line) PDF. The x-axis of  the $V_{trans}$ is printed on the bottom, and x-axis of $V_{rad}$ PDF is on the top of the plot.}
\end{figure}

The results are displayed on Figure 16. The $V_k$-$M_{2i}$ joint probability distribution shows that the kick given to NS2 at birth is between 5 and 510 km s$^{-1}$, while the progenitor mass $M_{2i}$ is between 1.25 and 4.80 $M_\sun$ with 95\% confidence. The $\theta'$ and $\theta_t$ PDF 's are similar to those of PSR J1756-2251. Both polar and planar kicks are possible, with the planar kicks having a higher probability. The polar angle $\theta_t$ is less than 51.5$^\circ$ with 95\% confidence. At the measured distance, the total transverse velocity of PSR J1906+0746 at the current time is between 21 and 376 km s$^{-1}$, and the radial velocity is between $-176$ and 297 km s$^{-1}$ with 95\% confidence. It is most likely that this binary is moving away from us. Furthermore, right before its supernova, the progenitor of NS2 must have been overflowing its Roche lobe, transferring mass to NS1.

Our limits on $V_k$ and $M_{2i}$ are tighter constrained than Wang et al. (2006), because they did not have a detailed analysis on this system. Instead, they calculated the allowed $V_k$ with an assumption that $M_{2i}$ was between 2.1 and 8.0 M$_\sun$.

\section{Discussion}

The anisotropic mass and neutrino ejection in a supernova explosion gives a recoil natal kick to the newly formed neutron star. Iron core collapse supernova and ECS produce natal kicks of different strength because of the different physical conditions of the nascent neutron star immediately before the supernovae.

There are several reviews (Bethe 1990, Mezzacappa 2005, Woosley \& Bloom 2006, Kotake et al 2006, and Janka et al. 2007) that summarize our current understanding of the iron core collapse supernova. After iron is synthesized in the core, the massive star reaches the final stage of hydrostatic nuclear burning, because the synthesis of any heavier elements requires energy instead of releasing energy. When the iron core grows by silicon shell burning around the core to a mass above the Chandrasekhar mass limit, electron degeneracy pressure cannot support the core any more and it starts to collapse. The iron core continues to collapse until it reaches nuclear density (i.e., $\approx 10^{14}$ g cm$^{-3}$). Since nuclear matter has a much lower compressibility, the collapse decelerates and the core bounces back because of the increased nuclear matter pressure. This drives a shock wave and eventually explodes the outer layers of the massive star away in a supernova explosion. 

On the other hand, the ECS scenario is described in the reviews of Miyaji et al. (1980) and Nomoto (1987). After carbon burning within the core of a post main sequence massive star, a $O - Ne - Mg$ core is formed. Neon cannot be ignited if the mass of the $O - Ne - Mg$ core is less than the critical mass of 1.37 M$_\sun$ for neon ignition. As the $O - Ne - Mg$ core is cooled by neutrino emissions, it becomes strongly degenerate. When the mass of the core grows by carbon shell burning and approaches the Chandrasekhar mass limit, electron captures onto $^{24}Mg$ and $^{24}Na$ take place. This leads to a decrease in the electron degeneracy pressure and the Chandrasekhar mass limit, which induces a rapid contraction of the core. The rapid core contraction ignites the oxygen deflagration, which incinerates materials into nuclear statistical equilibrium (NSE). Due to the rapid electron capture onto NSE elements, the collapse of the core accelerates. Similar to the iron core collapse picture, when the core reaches the nuclear matter density, it bounces back and drives a shock wave that eventually explodes the star in a supernova. Guti\'errez et al. (2005) show that the abundance of $^{24}Mg$ in $O - Ne - Mg$ core needs to be greater than 15\% , in order to have NSE developed through electron captures. However, in simulations of evolving massive AGB stars by Siess (2007), the abundance of $^{24}Mg$ after carbon burning is smaller by a factor of $\sim$10 than what is required to drive an explosion by electron captures. Therefore, the mechanism of developing an ECS is not a settled issue yet. In addition, Podsiadlowski et al. (2004) studied the binary evolution and the dynamics of core collapse, and suggested that ECS could only occur in interacting binaries, but not in single stars.

In recent simulations of supernova explosions from the collapse of $O - Ne - Mg$ cores by Kitaura et al. (2006), they found these supernovae are powered by neutrino heating and neutrino-driven wind of the nascent neutron star. Scheck et al. (2006) showed that the shock wave in these supernovae can propagate outwards on a relatively short timescale after the rebound of the core, which means the non-radial hydrodynamics instabilities do not have time to merge and grow to global asymmetry before the anisotropic pattern freezes out in the accelerating outward motion of the shock wave. As a result, \em \it the natal kick of an ECS due to anisotropic mass ejection is expected to be fairly small\rm \,(see Scheck et al. 2004). 
Furthermore, as the $O - Ne - Mg$ core that eventually explodes in an ECS is not massive enough to ignite neon, \em \it the mass of the neutron star progenitor should be less than the progenitors that explode in an iron core collapse supernova\rm. 

The present analysis is the first to include \em all \em currently known DNS systems and account of all orbital and kinematic constraints. We also employ a novel method for dealing with the uncertainty due to the un-measurable radial velocities and we focus on the derived PDFs and associated confidence levels, instead of just the most likely values, which can be misleading at times. Our results are summarized in Table 2; the derived constraints are consistent with earlier studies (when available) but typically limit parameters to narrower ranges,  and in this sense they represent the best available constraints on the formation of neutron stars in DNS systems.

In the context of our current understanding of massive star core collapse, we can use our results to draw a number of conclusions:
\\(1) PSR J0737-3039 has a $V_k$ upper limit of 120 km/s and a $M_{2i}$ upper limit of 1.9 $M_\sun$ at 95\% confidence. Therefore, the formation of NS2 (i.e., pulsar B) likely occurred through an ECS event. This is consistent with the speculation of Podsiadlowski et al. (2005) based on the space velocity and orbital eccentricity of PSR J0737-3039. For PSR B1534+12 and PSR B1913+16, $V_k$ are 150 - 270 km/s and 190 - 450 km/s (95\%) respectively, which means NS2 in both system must have received a significant recoil natal kick at birth. PSR B1534+12 has a $M_{2i}$ upper limit of 3.4 $M_\sun$ (95\%), while PSR B1913+16 has a $M_{2i}$ upper limit of 5.0 $M_{2i}$ (95\%). Because of the relatively high $V_k$ ranges and $M_{2i}$ upper limits, NS2 in both systems are probably formed through an iron core collapse supernova event. 
\\(2) PSR J1518+4904 has a $V_K$ upper limit of 110 km/s and a $M_{2i}$ upper limit of 4.7 $M_\sun$ at 95\% confidence. Even though NS2 likely received a low recoil velocity at birth, we cannot firmly conclude which type of supernova occurred during the formation of NS2, due to the relatively high $M_{2i}$ upper limit. 
\\(3) PSR J1756-2251 has a $V_k$ upper limit of 80 km/s and a $M_{2i}$ upper limit of 1.90$M_\sun$ at 60\% confidence, hence the formation of NS2 might possibly relate to an ECS event. However, the $V_k$ and $M_{2i}$ upper limits are 185 km/s and 2.65 $M_\sun$ respectively at 95\% confidence, which means the likelihood of NS2 formed through an ECS event is lower than that of PSR J0737-3039.
\\(4) PSR J1811-1736, PSR J1829+2456, and PSR J1906+0746 have $V_k$ upper limits of several hundreds of km/s, and $M_{2i}$ upper limits ranging from 4.80 to 8.0 $M_\sun$ (i.e., the conventional limit on neutron star progenitor mass) at 95\% confidence. Since both low and high $V_k$ and $M_{2i}$ are possible, we cannot conclude which type of supernova event is more favorable for the formation history of NS2 in these systems.
\\(5) For all of the known DNS except PSR B1534+12, all available constraints are consistent with imparting a polar or planar kick to NS2 at birth.  For PSR B1534+12, the kick direction is constrained to be polar.
\\(6) For PSR B1534+12, PSR J0737-3039 and PSR J1906+0746, the pre-supernova orbit is so tight that they cannot avoid a RLO from NS2 progenitor to NS1 just before the second supernova explosion.
\\(7) Furthermore, despite the low orbital eccentricity in PSR J1829+2456 and PSR J1906+746, NS2 could have a high progenitor mass and have received a high recoil natal kick at birth, as shown in the 2D joint PDF of these systems (see Figures 15, and 16).
\\(8) We also tested the dependence of our PSR J0737-3039 results on the current upper limit of the spin-orbital misalignment angle $\theta_t$ of NS1 (i.e., pulsar A). As shown in Figure 8, the $\theta_t$ PDF peaks strongly at $\sim 0^\circ$, so choosing an upper limit of 15$^\circ$ or 60$^\circ$ does not affect our results noticably.

\acknowledgments
We thank Rachel Dewey and Ingrid Stairs for useful discussions. This work is supported by a NSF CAREER grant AST-0449558 awarded to Vassiliki Kalogera.

\end{document}